\documentclass{article}
\usepackage{arxiv}
\usepackage[T1]{fontenc}
\usepackage{url}

\usepackage{graphicx}
\usepackage{amsmath} 
\usepackage{amssymb}
\usepackage{booktabs}   
\usepackage{array}
\usepackage{tabularx}
\usepackage{subfigure}

\usepackage{authblk}

\title{LSA: A Long-Short-term Aspect Interest Transformer for Aspect-Based Recommendation}

\author[1]{Le Liu}
\author[2]{Junrui Liu\thanks{Corresponding author}}
\author[1]{Yunhan Gao}
\author[1]{Ziheng Wang}
\author[3]{Tong Li}

\affil[1]{Beijing-Dublin International College, Beijing University of Technology, Beijing, China}
\affil[2]{Beijing Institute of Graphic Communication, Beijing, China}
\affil[3]{College of Computer Science, Beijing University of Technology, Beijing, China}

\renewcommand{\headeright}{Preprint}
\renewcommand{\shorttitle}{LSA}
\date{}

\begin{document}
\maketitle            

\begin{abstract}
Aspect-based recommendation methods extract aspect terms from reviews, such as price, to model fine-grained user preferences on items, making them a critical approach in personalized recommender systems.
Existing methods utilize graphs to represent the relationships among users, items, and aspect terms, modeling user preferences based on graph neural networks. 
However, they overlook the dynamic nature of user interests - users may temporarily focus on aspects they previously paid little attention to - making it difficult to assign accurate weights to aspect terms for each user-item interaction.
In this paper, we propose a long-short-term aspect interest Transformer (LSA) for aspect-based recommendation, which effectively captures the dynamic nature of user preferences by integrating both long-term and short-term aspect interests.
Specifically, the short-term interests model the temporal changes in the importance of recently interacted aspect terms, while the long-term interests consider global behavioral patterns, including aspects that users have not interacted with recently.
Finally, LSA combines long- and short-term interests to evaluate the importance of aspects within the union of user and item aspect neighbors, therefore accurately assigns aspect weights for each user-item interaction.
Experiments conducted on four real-world datasets demonstrate that LSA improves MSE by $2.55\%$ on average over the best baseline.
% The source code is available at \url{www.baidu.com}

\end{abstract}

\section{Introduction}
Aspect-based recommendation methods effectively learn the fine-grained user preferences and have attracted significant attention~\cite{DBLP:conf/cikm/ChinZJC18/ANR,DBLP:journals/kbs/QiuGCG16,DBLP:journals/ijon/YangYLNW16}.
Existing approaches primarily employ two paradigms: CNN-based and GNN-based methods.
CNN-based methods~\cite{DBLP:journals/mlc/JoseMA24,ARM} leverage CNN to extract aspect features from reviews.
In contrast, recent graph-based methods~\cite{DBLP:conf/wsdm/LiuLWTFY25,DBLP:journals/tkde/LiuYZMNZ23} explicitly model user-item-aspect relations using graph structures to better capture contextual and sentiment-aware representations.

However, existing methods only focus on modeling static user preferences from historical reviews and overlook the dynamic nature of user interests.
Users may temporarily focus on aspects they previously paid little attention to, which makes themn difficult to assign accurate weights to aspect terms for each user-item interaction~\cite{DBLP:conf/kdd/Bauman0T17,DBLP:conf/wsdm/LiuLWTFY25,DBLP:conf/aaai/OuZZLLA25}.
For example, a user expresses consistent interest in aspect terms, such as ``sound quality'' and ``build quality'' in history reviews, showing a long-term preference for high-performance audio equipment.
The user then often mentioned ``portability'' in recent product reviews, suggesting a short-term interest in ``portability'' due to a possible upcoming trip. 
Long-term interests represent the stable interests of users, while short-term interests are the preferences of users recently.

In this paper, we propose a Long-Short-term Aspect interest Transformer(LSA) for aspect-based recommendation, which effectively captures the dynamic nature of user preferences by integrating both long-term and short-term aspect interests.
% 图构建
Specifically, to capture users' transient needs, we propose a time-aware Transformer that models temporal changes in the importance of recently interacted aspect terms.
% 长期兴趣
Furthermore, LSA models the long-term interests by considering global behavioral patterns, including aspects that users have not interacted with recently.
Those two interests are integrated through a gated fusion mechanism.
To generate a personalized prediction, we introduce an interest-aware aspect term aggregation method, which uses user preferences to justify the importance of aspect terms, over the union of user- and item-specific aspect terms.
% FM
The resulting aspect representation is concatenated with user and item embeddings and passed into a factorization machine for final scoring.

The main contributions of this paper are summarized as follows:(1) We propose a Long-Short-term Aspect Interest Transformer(LSA) for aspect-based recommendation, which models the evolution of user interests through considering long- and short-term of interests.
(2) We design a dual aspect selection method to capture users' long-term and short-term interests effectively at the aspect granularity.
(3) We introduce an interest-aware aspect term aggregation mechanism to align user preferences with item attributes for personalized prediction.
(4) Extensive experiments conducted on four real-world datasets demonstrate that LSA improves MSE by an average of $2.55\%$ over the best baseline.

\section{Related Work}
In this section, we review previous work relevant to our work, including Aspect-Based Recommendation and Modeling Long- and Short-Term User Interests.

\subsection{Aspect-Based Recommendation}
Aspect-based recommender systems leverage fine-grained information from user reviews (i.e. aspect–sentiment pairs) to provide more personalized recommendations. 
Early works start to explore the sentiment of each aspects in user review instead of serving as a single source to explore the user's preference on different aspects~\cite{DBLP:journals/kbs/QiuGCG16,DBLP:journals/ijon/YangYLNW16}. 
Building on this idea, researchers have proposed models to extract aspect terms from reviews and incorporate them into recommendation algorithms.
For example, ANR\cite{DBLP:conf/cikm/ChinZJC18/ANR} applies attention-based neural networks to identify aspect-level semantics and employs co-attention to capture asymmetric importance of aspects between users and items. 
Subsequent works integrate sentiment reasoning into aspect modeling.
ARPM~\cite{DBLP:journals/is/LaiH21} incorporates social signals for preference modeling, while CARP~\cite{DBLP:conf/sigir/LiQPQDW19} introduces logic-unit based capsule reasoning for fine-grained aspect-sentiment understanding. SENGR~\cite{DBLP:journals/isci/ShiWGHCZH22} integrates sentiment-aware review attention and graph-based user-item-user modeling to enhance rating prediction. 
Most recently, APH~\cite{DBLP:conf/wsdm/LiuLWTFY25} models the aspect performance of items using a sentiment-aware hypergraph neural network, introducing a novel aspect hypergraph to aggregate conflicting user sentiments for fine-grained recommendation.

\subsection{Modeling Long and Short-Term User Interests}
Users' preferences are not static. 
Traditional models mainly consider long and short-term user interests at the item level. 
PLASTIC~\cite{DBLP:conf/ijcai/ZhaoWYGYC18} proposes an adversarial framework that fuses long‑term matrix‑factorization profiles with short‑term RNN session dynamics. 
Yu et al.~\cite{DBLP:conf/ijcai/YuLML019} model short‑term intent with an LSTM equipped with time‑ and content‑aware controllers, then adaptively fuse it—via attention—with an attentive long‑term preference vector.
Most recently, LS-TGNN~\cite{DBLP:conf/aaai/OuZZLLA25} leverages a temporal graph neural network and self-supervised contrastive learning to disentangle and model users' evolving long- and short-term interests, significantly enhancing session-based recommendation performance.

\section{Methodology}

\begin{figure}[!htbp]
\centering
\includegraphics[width=1\linewidth]{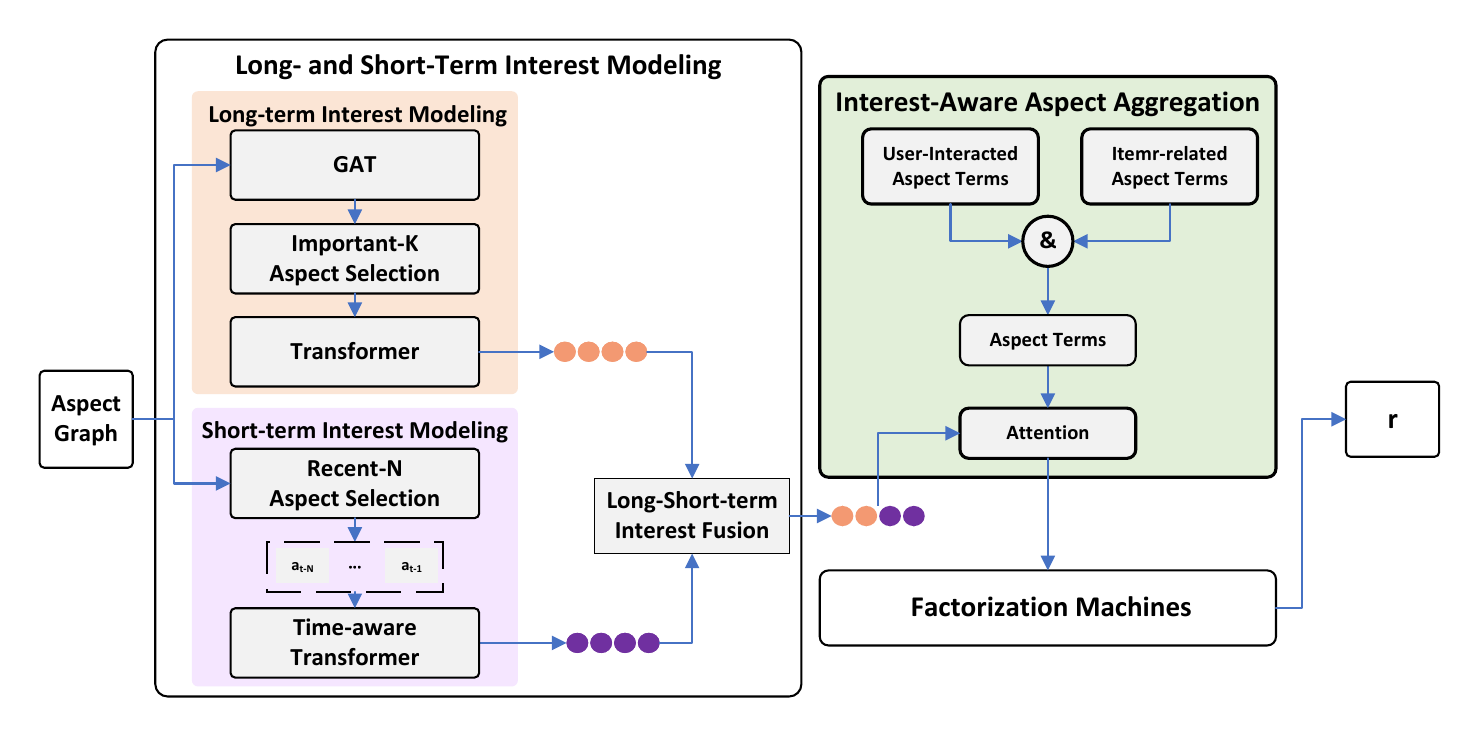}
\vspace{-1cm}
\caption{Architecture of LSA.}
\label{fig:wide}
\vspace{-0.5cm}
\end{figure}

To model the dynamic nature of user interests at the aspect level, this paper proposes LSA for aspect-based recommendation, with its architecture illustrated in Figure~\ref{fig:wide}. 
Specifically, LSA (1)~extracts aspect terms from reviews and constructs a \emph{user–item–aspect graph};  (2)~captures long‑term preferences by a graph Transformer; (3)~captures short‑term preferences by forming time‑ordered aspect sequences within temporal window; and (4)~fuses the two views through a long–short attention mechanism to aggregate aspect terms. Finally, we describe the final prediction layer.

\vspace{-0.5cm}

\subsection{Aspect Graph}
We initially extract aspect terms from reviews to construct the aspect graph.
Following existing methods~\cite{DBLP:conf/emnlp/LiCKHCW21,DBLP:conf/wsdm/LiuLWTFY25}, we choose a rule-based unsupervised method, which considers three dependency relations to extract aspect terms: \textit{amod}, \textit{dobj} and \textit{nsubj+acomp}.
After extracting aspect-sentiment from each review, we construct user-item-aspect graph, where nodes represent users, items, and aspect terms.
Edges are created between nodes based on review interactions, and edge weights reflect either aspect frequency or user-item rating.
In addition, the interaction timestamps are also recorded to support time-aware modeling.

\subsection{Long- and Short-Term Interest Modeling}

Users’ interests are shaped by both stable long-term behavior and transient short-term needs.
Thus, LSA models short-term interests using a time-aware Transformer that encodes the most recent $N$ aspect terms within a $T$-day window, and long-term interests via a graph Transformer over the user-item-aspect graph with an important-K aspect selection strategy to filter out irrelevant or temporary aspects

\subsubsection{Modeling Long-Term Interest}

Users’ long-term interests reflect stable preferences formed over extended periods, but existing GNN-based methods overlook future-relevant aspects by only aggregating interacted aspect nodes.
To address this, we employ a graph Transformer that leverages self-attention to capture both global and local structural dependencies in the user-item-aspect graph, while an Important-K Aspect Selection method filters out noisy aspects by retaining only the most representative ones based on interaction frequency and rating preferences, ensuring more efficient and focused long-term interest modeling.

First, \textbf{Edge Weight Encoding} quantifies the frequency of interactions between users/items and aspects, under the assumption that higher interaction frequency typically signals stronger and more stable preferences. 
The interaction score is computed as:
\begin{equation}
s_{u,a}^{\text{edge}} = \sigma\left( \exp\left( \frac{w_{u,a}}{100} \right) \right)
\end{equation}
where $w_{u,a}$ represents the frequency of interactions between user $u$ and aspect $a$, and $\sigma(\cdot)$ denotes a sigmoid function for normalization.
The denominator 100 is an empirically chosen scaling factor to prevent the exponential term from becoming excessively large.

In addition to interaction frequency, users' interest in certain aspects can often be inferred from ratings. 
Thus, \textbf{Rating Preference Encoding} incorporates semantic correlations derived from user ratings, reflecting user sentiment and perceived importance of specific aspects. The rating-based semantic relevance score is computed as:
\begin{equation}
s^{\text{pre}}_{u,a} = f(\mathbf{W}_1 \mathbf{y}_u, \mathbf{e}_a),
\end{equation}
where $\mathbf{y}_u$ is rating preference embedding of user $u$, $\mathbf{W}_1$ is a learnable transformation matrix, and $\mathbf{e}_a$ represents the embedding of aspect $a$ obtained from the graph encoder.
The function $f(\cdot)$ denotes a similarity function that measures the correlation between the transformed user embedding and the aspect embedding.

By summing these two encoding scores, we derive a comprehensive relevance score for user $u$ to rank entire aspects:
\begin{equation}
s_{u,a} = s_{u,a}^{\text{edge}} + s_{u,a}^{\text{pre}}.
\end{equation}
The user node is then added to the beginning of this sequence, forming the input sequence for the Transformer for modeling long-term interest as $\mathcal{A}_u^l = [\mathbf{e}_u, \mathbf{e}_{a_1}, \dots, \mathbf{e}_{a_K}]$.

In practice, we enhance the Transformer by replacing its self-attention mechanism with GATv2~\cite{GATv2}.
After the encoding of the Transformer, we obtain the user's long-term interest representation, denoted as $\hat{\mathbf{e}}_u^l$. This representation encapsulates both structural insights and semantic preferences, precisely capturing the user's stable and enduring aspect interests.

\subsubsection{Modeling Short-Term Interest}

Users’ interests evolve over time, and recent interactions reveal their short-term preferences—for instance, a guitarist’s reviews may shift from beginner-focused features to performance-level aspects over a year.
To capture this dynamic behavior, we propose a Recent-$N$ Aspect Selection method that selects the most recent $N$ aspect terms from a user’s history, highlighting their current short-term interests.

Formally, given a sequence of user interactions, each associated with aspect terms and timestamps, we first define the user's interaction history as $ H_u = \{(a_i, t_i) | t_i < t_{\text{curr}}, 1 \leq i \leq n\}$,
where $a_i$ denotes an interacted aspect term, and $t_i$ is the timestamp of interaction. From this historical set, we filter and retain valid interactions within a specified temporal window, ensuring temporal relevance.
Next, we sort the valid interactions in descending order by timestamp and select the most recent $N$ aspect to form a short-term sequence, thereby emphasizing the recency of the interactions and their direct relevance to current preferences.
We then concatenate the user node with the embeddings of these recent aspects, constructing an ordered sequence $\mathcal{A}_u^s = [\mathbf{e}_u, \mathbf{e}_{a_1^{(t)}}, \dots, \mathbf{e}_{a_N^{(t)}}]$ as input to the Transformer encoder.

After processing with the Transformer encoder, the embedding corresponding to the user node (i.e., the first token in the sequence) is extracted as the user's refined short-term interest representation, denoted as $\hat{e}_u^s$.

\subsubsection{Gated Long-Short-term Interest Fusion}
This layer attempts to combine both long- and short-term interest via a gate-residual mechanism. 
Firstly, it projects long-term and short-term representations into a shared latent space:
\begin{align}
\mathbf{e}_{u}^{\,l'} &= \mathbf{W}_{l}\,\hat{\mathbf{e}}_{u}^{\,l},
~~~~
% \\[2pt]
\mathbf{e}_{u}^{\,s'}= \mathbf{W}_{s}\,\hat{\mathbf{e}}_{u}^{\,s},
\end{align}
\noindent%
where $\mathbf{W}_{l},\mathbf{W}_{s}\in\mathbb{R}^{d\times d}$ are learnable linear projections.
Next, a gated neural network dynamically balances contributions from long-term and short-term embeddings:
\begin{align}
\mathbf{g}_{u} \;=&\;
      \sigma\!\Bigl(
            \mathbf{W}_{g}
            \bigl[\,
               \mathbf{e}_{u}^{\,l'} \,\Vert\, \mathbf{e}_{u}^{\,s'}
            \bigr]
            + \mathbf{b}_{g}
      \Bigr),\\[4pt]
\mathbf{e}_{u} \;=&\;
      \mathbf{g}_{u} \odot \mathbf{e}_{u}^{\,l'}
      \;+\;
      (1-\mathbf{g}_{u}) \odot \mathbf{e}_{u}^{\,s'}
      \;+\;
      \mathbf{e}_{u}^{\,l'}.
\end{align}
where $\sigma(\cdot)$ denotes the sigmoid activation function, and $\odot$ represents multiplication.

Finally, we concatenate the non-linear transformation of the user embedding $\textbf{y}_u$ after passing it through an MLP layer with the aspect-aware aggregation representation to generate the final representation of the user $u$ as follows,
\begin{align}
\mathbf{p}_{u} \;=&\;
      \operatorname{ReLU}\!\bigl(
         \mathbf{y}_{u}\mathbf{W}_{3} + \mathbf{b}
      \bigr)\oplus\; \mathbf{e}_{u}.
\end{align}
Similarly, we can get the representation of the item $i$ as $\mathbf{q}_i$.

\subsection{Interest-Aware Aspect Term Aggregation}

Aspect terms influence user preferences and item features differently, but existing methods often predict ratings using only user and item representations.
To capture both known and novel influences, we construct the aspect term set as the union of user and item aspects, allowing the model to consider previously unobserved aspects that may affect the interaction.
Formally, for user-item pair $(u, i)$, we define the candidate aspect set as:
\begin{equation}
    \mathcal{A}_{u,i} = \mathcal{A}_u \cup \mathcal{A}_i
\end{equation}
where $\mathcal{A}_u$ and $\mathcal{A}_i$ denote the aspect terms extracted from the user's historical reviews and from the candidate item's associated aspects, respectively.

To model the unequal contribution of different aspect terms in a given interaction, we adopt an attention-based aggregation mechanism considered on the user-item context.
Firstly, the information of the user and the item is combined together by a fusion layer:
\begin{equation}
    \mathbf{f}_{ui} = \text{ReLU}(\mathbf{W}_f \cdot [\mathbf{p}_u \oplus \mathbf{q}_i] + \mathbf{b}_f)
\end{equation}
where $\mathbf{p}_u$ is the fused user interest vector, $\mathbf{q}_i$ is the item representation, and $\mathbf{W}_f$, $\mathbf{b}_f$ are learnable parameters. The fused representation $\mathbf{f}_{ui}$ captures the contextual interaction signal.

We then use this fused vector as a query to attend over the candidate aspect terms in $\mathcal{A}_{u,i}$. Each aspect term $a_k \in \mathcal{A}_{u,i}$ has been embedded as $\mathbf{e}_{a_k} \in \mathbb{R}^d$, and is projected into key–value pairs using learnable transformations.
The user--item fusion representation $\mathbf{f}_{ui}$ is projected into query space to capture interaction-specific context:
\begin{align}
    \mathbf{q}_{ui} = \mathbf{W}_q \mathbf{f}_{ui},\quad
    \mathbf{k}_k = \mathbf{W}_K \mathbf{e}_{a_k},\quad \mathbf{v}_k = \mathbf{W}_V \mathbf{e}_{a_k}
\end{align}

Finally, the final aspect representation is obtained by aggregating different aspect terms according to their attention scores.
\begin{equation}
\mathbf{h}_a = \sum_{k=1}^{|\mathcal{A}_{u,i}|} {\frac{\exp(\mathbf{q}_{ui}^\top \mathbf{k}_k)}{\sum_{j=1}^{|\mathcal{A}_{u,i}|} \exp(\mathbf{q}_{ui}^\top \mathbf{k}_j)}} \cdot \mathbf{v}_k
\end{equation}

\subsection{Prediction}
This section adopts an Factorization Machines (FM) layer to predict final scores \cite{RN3}. 
It captures higher-order interactions between users, items, and aspects. 
Specifically, we concatenate the user, item, and aspect representation:
\begin{equation}
\mathbf{x} = \mathbf{p}_u \oplus \mathbf{q}_i \oplus \lambda \cdot \mathbf{h}_a
\end{equation}
The prediction score $\hat{r}_{ui}$ is computed as follows
\begin{equation}
\hat{r}_{ui} = b_0 + b_u + b_i+ \mathbf{x} \mathbf{w}^\top  + \sum_{i=1}^{d^{\prime}} \sum_{j=i+1}^{d^{\prime}}<\mathbf{v}_i, \mathbf{v}_j>\mathbf{x}_i \mathbf{x}_j ,
\end{equation}
where $b_0$, $b_u$, and $b_i$ are the global, user, and item biases, respectively. $\mathbf{w} \in \mathbb{R}^{1 \times d'}$ is the coefficient vector of the linear term. $\langle \mathbf{v}_i, \mathbf{v}_j \rangle$ denotes the inner product of the latent vectors associated with the $i$-th and $j$-th feature dimensions. $x_i$ is the value of the $i-$th dimension of $x$.
The overall objective is to minimize the Mean Squared Error (MSE) between predicted and true ratings over the training set.

\section{Experiment}
In this section, we complete experiments to identify the performance of the proposed method. 
We first describe the experiment setup, then show the comparison results with diverse baselines, and finally explore the effect of model components.

\subsection{Experimental Setup}

\subsubsection{Dataset}
Experiments are done on the Amazon review datasets\footnote{\url{http://jmcauley.ucsd.edu/data/amazon/}}, which have been widely used in recommendation research. 
Following existing works~\cite{DBLP:journals/tkde/LiuYZMNZ23}, we use 4 core subsets: Musical Instrument, Office Product, Beauty, and Video Games (respectively represented by Music, Office, Beauty, and Games).

\subsubsection{Baseline Methods}
We compared our method with the following SOTA baselines: review-based method includes DeepCoNN~\cite{BL6}, NARRE~\cite{BL7}, CARL~\cite{BL8}, DAML~\cite{BL9}, DSRLN~\cite{BL10}; and a graph-based method RGNN~\cite{DBLP:journals/tkde/LiuYZMNZ23}.

\subsubsection{Evaluation Metrics} 
To evaluate the performance of our method, we adopt three metrics, i.e., Mean Square Error (MSE), Mean Absolute Error (MAE) and Normalized Discounted Cumulative Gain (NDCG)~\cite{DBLP:journals/tkde/LiuYZMNZ23,BL8,BL6}. These metrics are
commonly used in existing recommendation studies, and we follow this practice
to ensure fair comparability with prior work.
For each experiment, we randomly choose 20\% of the user-item review pairs (denoted by $D_{test}$) for evaluating the model performance in the testing phase, and the remaining 80\% of the review pairs (denoted by $D_{train}$) are used in the training phase.

\subsection{Performance Comparison}
Table~\ref{tab:evaluation} summarizes the performance comparison between our proposed method and several baselines across four datasets. 
The best results are highlighted in boldface, and the second-best are underlined. 

LSA achieves the lowest MSE on every dataset, with relative improvements over the strongest baseline ranging from 0.41\% (Games) to 5.52\% (Music) -- an average MSE gain of about 2.55\%. 
In terms of MAE, LSA also demonstrates strong performance on most datasets, particularly Music, Office, and Games. On Beauty, while DSRLN slightly outperforms LSA, the overall performance remains competitive.
For NDCG@10(N@10), LSA achieves the best performance on three datasets: Music, Office, and Beauty. 
However, on the Games dataset, LSA shows slightly lower results compared to DAML and DSRLN.
One contributing factor could be the model's frequency-based long-term aspect selection, which might emphasize common aspect terms while under-representing niche user preferences.
Furthermore, we does not optimize for ranking objectives, it may be result poor performance on top-k recommendation metrics lisk N@10.
Overall, LSA consistently delivers robust improvements across diverse datasets and metrics, affirming the effectiveness of integrating fine-grained aspect terms with long- and short-term user preference modeling. 

\begin{table}[htbp]
\small
\centering
\caption{Comparison results. \textbf{Bold} values indicate the best performance, and \underline{underlined} values indicate the second best.}
\label{tab:evaluation}
\scalebox{0.9}{
\begin{tabular}{lcccccccccccc}
\toprule
\textbf{Model} & \multicolumn{3}{c}{\textbf{Music}} & \multicolumn{3}{c}{\textbf{Office}} & \multicolumn{3}{c}{\textbf{Beauty}} & \multicolumn{3}{c}{\textbf{Games}} \\
         & MSE    & MAE    & N@10  & MSE    & MAE   & N@10 & MSE    &  MAE   & N@10 &  MSE   & MAE    & N@10 \\
\midrule
DCN      & 0.7909 & 0.7590 & 0.977   & 0.7315 & 0.7113 & 0.973   & 1.2210 & 0.9209 & 0.966   & 1.1234 & 0.8750 & 0.971  \\
NARRE    & 0.7688 & 0.6950 & 0.978   & 0.7266 & 0.6812 & 0.976   & 1.1997 & 0.9228 & 0.971   & 1.1120 & 0.7990 & 0.968 \\
CARL     & 0.7632 & 0.6771 & 0.980   & 0.7193 & 0.6477 & 0.978   & 1.2250 & 0.8809 & 0.966   & 1.1308 & 0.7980 & 0.969\\
DAML     & 0.7401 & 0.6512 & \underline{0.982}   & 0.7164 &\underline{0.6122}& 0.978& 1.2175 & 0.9192 & 0.967& 1.1086 & 0.7880 & \textbf{0.979}\\
DSRLN    & 0.7538 & 0.6207 & 0.781   & 0.7131 & 0.6210 & 0.974   & 1.1951 & \textbf{0.8105} &0.967 & 1.1205 & 0.7713 & \textbf{0.979} \\
\midrule
RGNN     & \underline{0.7319} &\underline{0.6019}&  \underline{0.982} & \underline{0.7125}&0.6231& 0.983     & \underline{1.1885} & 0.9303 &  0.973& \underline{1.0996} & \underline{0.7864} & 0.976\\
\midrule
\textbf{LSA} & \textbf{0.6915} & \textbf{0.5800}& \textbf{0.986}   & \textbf{0.6897} & \textbf{0.6099} &  \textbf{0.985}  & \textbf{1.1760} & \underline{0.8119}& \textbf{0.974}    & \textbf{1.0950} & \textbf{0.7799} & \underline{0.977}\\
\bottomrule
\end{tabular}
}
\vspace{-0.5cm}
\end{table}

\subsection{Ablation study}

This experiment develops some variants by removing aspect attnetion, fusion, short-term and long-term layers to verify their contribution for LSA.
The result of ablation experiments is shown in Table~\ref{tab:ablation_mse}. 
LSA outperforms all variants, demonstrating that each module contributes positively to performance. 
A user has mentioned multiple aspects, but not all are equally relevant to a specific interaction. 
Therefore, we design an interest-aware aspect aggregation layer that attends to informative aspects based on user-item context.
If we replace the gated fusion with a simple average, it will limit the model's ability to adaptively weigh long- and short-term interests, resulting in poor performance.

Overall, this ablation experiment confirms the effectiveness and complementary contributions of the aspect-aware attention mechanism, gated fusion strategy, and distinct long- and short-term interest encoders in our method.

\begin{table}[htbp]
\centering
\caption{MSE results of ablation study.}
\label{tab:ablation_mse}
\begin{tabular}{>{\centering\arraybackslash}p{4.2cm} 
                >{\centering\arraybackslash}p{1.2cm} 
                >{\centering\arraybackslash}p{1.2cm} 
                >{\centering\arraybackslash}p{1.2cm} 
                >{\centering\arraybackslash}p{1.2cm}}
\toprule
\textbf{Model Variant} & \textbf{Music} & \textbf{Office} & \textbf{Beauty} & \textbf{Games} \\
\midrule
LSA            & \textbf{0.6915}& \textbf{0.6897}  &\textbf{1.1760}  & \textbf{1.0950}  \\
w/o Aspect Attention  & 0.7030  & 0.7018  & 1.1762  & 1.0966  \\
w/o Fusion            & 0.7034  & 0.6942  & 1.2059  & 1.1390   \\
w/o Short Term        & 0.6993  & 0.6956  & 1.2065  & 1.1381  \\
w/o Long Term         & 0.6932  & 0.6948  & 1.2063  & 1.1393   \\
\bottomrule
\end{tabular}
\vspace{-0.5cm}
\end{table}

\subsection{Parameter Sensitivity Analysis}

In this section, we conduct a sensitivity analysis on three key hyperparameters: the number of long-term aspects ($K$), the number of recent aspects ($N$), varying each while fixing the others.
For space limitation, we show the experiment results of the Music dataset with fixed batch size=16 and num layers=2 when analyzing $K$ and $N$.
On the Music dataset, we observe that increasing $K$ generally improves performance, with the best MSE at $K=80$, likely due to richer user preference coverage. For $N$, the optimal value is 20, suggesting that a moderate temporal window best captures short-term intent.

\begin{figure}[ht]
\centering
\subfigure[The impact of $N$ (K=64)]{
    \includegraphics[width=0.4\textwidth]{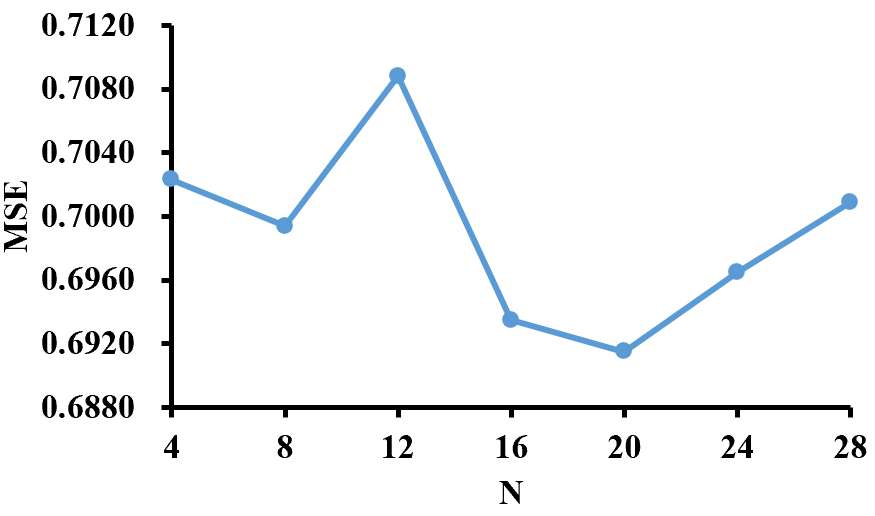}
    \label{a}}
% \hspace{1cm} % 减少水平间距
\subfigure[The impact of $K$ (N=20)]{
    \includegraphics[width=0.4\textwidth]{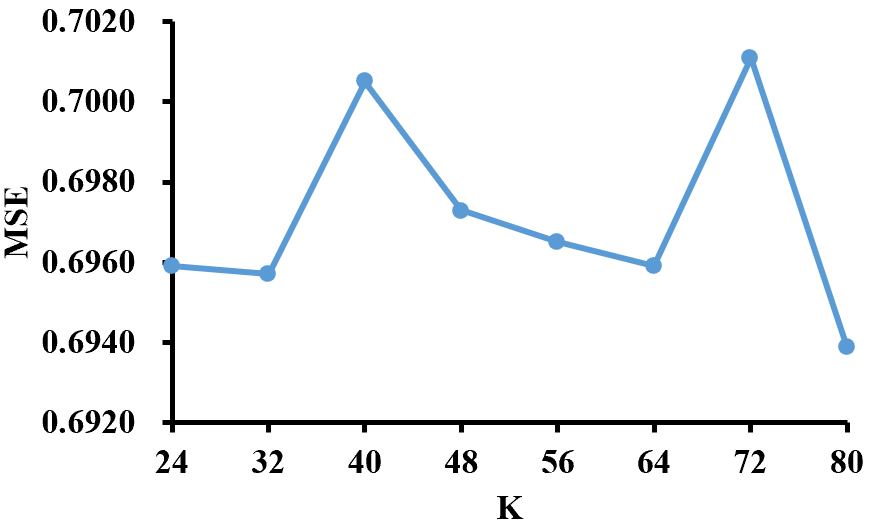}
    \label{b}}
\caption{MSE under different values of $K$ and $N$ on the Music dataset.}
\label{fig:3}
\vspace{-1cm}
\end{figure}

\section{Conclusion}
Existing aspect-based recommendation methods use GNN to model relationships among users, items, and aspect terms, which only focus on modeling static user preferences and overlook the dynamic nature of user interests
In this paper, we propose a long-short-term aspect interest Transformer for aspect-based recommendation, which considers long- and short-term user interests on aspect terms to learn the dynamic nature of user preferences on them. 
Experiments on four real-world datasets demonstrate the effectiveness of LSA. 
In future work, we plan to explore sentiment-aware aspect modeling to enhance the representation, aiming for more personalized and interpretable recommendations. 

\section*{Acknowledgments}
This work was supported by the Beijing University of Technology Education and Teaching Research Project (No.ER2024KCA01),
the open project of Key Laboratory Ministry of Industry and Information Technology (No.ZCQP2024000337),
the open project of IoT Standards and Application Key Laboratory of the Ministry of Industry and Information Technology (No.202407),
and the National Training Program of Innovation and Entrepreneurship for Undergraduates of China (Beijing University of Technology) under Grant No. GJDC2024018383.

% \newpage

\bibliographystyle{splncs04}
\bibliography{ref}

\end{document}